\documentstyle[12pt]{article}
\begin{document}
\hfill{CCUTH-96-01}\par
\vskip 0.5cm
\centerline{\large{{\bf Internal $W$-emission and $W$-exchange 
Contributions}}}\par
\centerline{\large{{\bf to $B\to D^{(*)}$ Decays}}}\par
\vskip 1.0cm
\centerline{Chung-Yi Wu$^1$, Tsung-Wen Yeh$^1$ and Hsiang-nan Li$^2$ }
\vskip 0.3cm
\centerline{$^1$Department of Physics, National Cheng-Kung University,}
\centerline{Tainan, Taiwan, R.O.C.}
\vskip 0.3cm
\centerline{$^2$Department of Physics, National Chung-Cheng University,}
\centerline{Chia-Yi, Taiwan, R.O.C.}
\vskip 1.0cm
\centerline{\today}
\vfill
\baselineskip=2.0\baselineskip
\centerline{\bf Abstract}
We evaluate external $W$-emission, internal $W$-emission and $W$-exchange 
contributions to nonleptonic $B\to D^{(*)}$ decays based on the 
perturbative QCD formalism including Sudakov effects, whose ratio is found 
to be $1:+0.2:0.03i$ at the amplitude level. We observe that the internal 
$W$-emission contribution is additive to the external $W$-emission
contribution, and that the $W$-exchange contribution is negligible and
mainly imaginary, its real part being at least one order of magnitude
smaller than the imaginary part. Our predictions are consistent
with the CLEO data and with those obtained by the Bauer-Stech-Wirbel
method.

\newpage
\centerline{\large \bf 1. Introduction}
\vskip 0.3cm

The conventional approach to nonleptonic $B$ meson decays is the
Bauer-Stech-Wirbel (BSW) method \cite{BSW}, in which two parameters
$a_1$ and $a_2$ are associated with the external and internal 
$W$-emission amplitudes, respectively, and are determined by experimental
data. $a_1$ and $a_2$ were originally the linear combination of the Wilson 
coefficients $c_1$ and $c_2$ in the effective Hamiltonian 
\begin{equation}
H_{\rm eff}=\frac{G_F}{\sqrt{2}}V_{cb}V_{ud}^*[c_1(\mu)({\bar d}u)
({\bar c}b)+c_2(\mu)({\bar c}u)({\bar d}b)]\;, 
\label{eff}
\end{equation}
written as $a_1=c_1+c_2/N_c$ and $a_2=c_2+c_1/N_c$, $N_c$
being the number of colors. $H_{\rm eff}$
was derived from the Hamiltonian 
\begin{equation}
H=\frac{G_F}{\sqrt{2}}V_{cb}V_{ud}^*({\bar d}u)({\bar c}b)\;,
\label{ham}
\end{equation}
with hard gluon corrections taken into account by renormalization group 
(RG) methods. Here $({\bar q}_iq_j)={\bar q}_i\gamma_\mu(1-\gamma_5)q_j$ 
represents the $V-A$ current, $G_F$ is the Fermi coupling constant, and 
$V_{cb}$ and $V_{ud}$ are the Cabibbo-Kobayashi-Maskawa (CKM) matrix 
elements. 

The hadronic form factors involved in nonleptonic $B$ meson decays 
are usually assumed to take, say, a monopole or dipole ansatz 
\cite{CT}, and thus the extraction of $a_1$ and $a_2$ is model dependent. 
It has been found that the ratio $a_2/a_1$ from an individual fit 
to the CLEO data of $B\to D^{(*)}\pi(\rho)$ \cite{A} varies significantly
\cite{CT}. It has even been shown that an allowed domain $(a_1,a_2)$ 
exists for the three classes of decays ${\bar B}^0\to D^{(*)+}$, 
${\bar B}^0\to D^{(*)0}$ and $B^-\to D^{(*)0}$, only when the
experimental errors are expanded to a large extent \cite{GKKP}.
On the other hand, a negative $a_2/a_1$ and a positive $a_2/a_1$ were
concluded from the data of charm and bottom decays \cite{BSW,KP}, 
respectively, and this subject remains controversial. 

In the BSW model nonspectator contributions from $W$-exchange 
(or annihilation) diagrams are not included. In previous studies such
contributions were assumed to be negligible, though a convincing
justification is not yet available. Hence, it has been a challenging 
task to evaluate $W$-exchange contributions to nonleptonic $B$ meson 
decays reliably.

Recently, we have shown that the perturbative QCD (PQCD) formalism, 
which contains the resummation of large radiative corrections,
is applicable to $B\to D$ decays in the fast recoil region of the 
$D$ meson \cite{L1}, and have employed this formalism to
investigate the spectator contributions from external $W$-emission 
diagrams to $B\to D^{(*)}$ decays \cite{WYL}. In our approach all
the relevant form factors can be evaluated explicitly without resort
to models. For neutral $B$ meson decays such as ${\bar B}^0\to 
D^{(*)+}\pi^-(\rho^-)$, there are additional contributions from 
$W$-exchange diagrams. For charged $B$ meson decays such as $B^-\to 
D^{(*)0}\pi^-(\rho^-)$, both external and internal $W$-emission diagrams 
contribute. 

In this letter we shall extend the PQCD formalism to the study of
internal $W$-emission and $W$-exchange diagrams, whose contributions 
are computed reliably. Our work helps clarify the 
ambiguity in the extraction of $a_1$ and $a_2$ in the BSW model. A PQCD 
analysis of these contributions based on the Brodsky-Lepage theory 
\cite{BL} has been performed in \cite{CM}. However, this approach was
criticized in \cite{LL}: The parametrization of the 
fractional momentum carried by the light valence quark in a $B$ meson
does not satisfy the on-shell requirement from the parton model, such
that extra imaginary contributions are introduced. Moreover, the 
predictions depend strongly on the model of meson wave functions 
\cite{CM,LL}.
\vskip 1.0cm

\centerline{\large \bf 2. Resummation}
\vskip 0.3cm

We review the investigation of the external $W$-emission diagrams as shown 
in Fig.~1(a), which involve six form factors $\xi_i$, $i=+$, 
$-$, $V$, $A_1$, $A_2$ and $A_3$, defined by the hadronic matrix 
elements of vector and axial vector currents \cite{WYL}. Assume that the 
momentum $P_1$ $(P_2)$, the mass $M_B$ ($M_{D^{(*)}}$) and the velocity
$v_1$ $(v_2)$ of the $B$ $(D^{(*)})$ meson are related by $P_1=M_Bv_1$
$(P_2=M_{D^{(*)}}v_2)$. The velocity transfer $\eta=v_1\cdot v_2$ 
is expressed in terms of the momentum transfer $q^2=(P_1-P_2)^2$ as
\begin{equation}
\eta=\frac{M_B^2+M_{D^{(*)}}^2-q^2}{2M_BM_{D^{(*)}}}\;.
\end{equation}
In the infinite mass limit of $M_B$ and $M_{D^{(*)}}$, the 
six form factors have the relations
\begin{equation}
\xi_+=\xi_V=\xi_{A_1}=\xi_{A_3}=\xi,\;\;\;\;  \xi_-=\xi_{A_2}=0.
\label{iwr}
\end{equation}
$\xi$ is the so-called Isgur-Wise (IW) function \cite{IW}, which is 
normalized to unity at zero recoil $\eta\to 1$ by heavy quark symmetry 
(HQS).

In the rest frame of the $B$ meson $P_1$ is written
as $P_1=(M_B/\sqrt{2})(1,1,{\bf 0}_T)$, and 
$P_2$ has the nonvanishing components \cite{L1}
\begin{eqnarray}
P_2^+=\frac{\eta+\sqrt{\eta^2-1}}{\sqrt{2}}M_{D^{(*)}}\;,
\;\;\;\;
P_2^-=\frac{\eta-\sqrt{\eta^2-1}}{\sqrt{2}}M_{D^{(*)}}\;.
\end{eqnarray}
Let $k_1$ ($k_2$) be the momentum of the light valence quark in the $B$ 
($D^{(*)}$) meson, satisfying $k_1^2\approx 0$ ($k_2^2\approx 0$). 
$k_1$ may have a large minus component $k_1^-$, defining the momentum 
fraction $x_1=k_1^-/P_1^-$, and small transverse components ${\bf k}_{1T}$. 
$k_2$ may have a large plus component $k_2^+$, defining $x_2=k_2^+/P_2^+$, 
and small ${\bf k}_{2T}$. As $\eta\to 1$ with $P_2^+=P_2^-=
M_{D^{(*)}}/\sqrt{2}$, the two meson wave functions 
strongly overlap, and the form factors $\xi_i$ are dominated by soft 
contributions. In the large $\eta$ limit with $P_2^+\gg M_{D^{(*)}}/
\sqrt{2}\gg P_2^-$, the $D^{(*)}$ meson behaves like a light meson
\cite{L1}. Then $B\to D^{(*)}$ transitions occur through hard gluon 
exchanges, to which PQCD is applicable. 

Double logarithms from radiative corrections to the $B$ ($D^{(*)}$) meson 
wave function $\phi_B$ ($\phi_{D^{(*)}}$) have been organized 
into an exponent $s$ using the resummation technique \cite{L1}.
We quote the results as 
\begin{eqnarray}
\phi_B(x_1,P_1,b_1,\mu)&=&\phi_B(x_1)\exp\left[-s(x_1P_1^-,b_1)-2
\int_{1/b_1}^{\mu}\frac{d{\bar \mu}}{\bar \mu}\gamma(\alpha_s({\bar \mu}))
\right]\;,
\nonumber \\
\phi_{D^{(*)}}(x_2,P_2,b_2,\mu)&=&\phi_{D^{(*)}}(x_2)\exp
\biggl[-s(x_2P_2^+,b_2)-s((1-x_2)P_2^+,b_2)
\nonumber \\
& &\left.-2\int_{1/b_2}^{\mu}
\frac{d{\bar \mu}}{\bar \mu}\gamma(\alpha_s({\bar \mu}))\right]\;,
\label{wp}
\end{eqnarray}
where $b_1$ ($b_2$) is the Fourier conjugate variable of $k_{1T}$ 
($k_{2T}$), and can be regarded as the spatial extent of the $B$ ($D^{(*)}$) 
meson. $\mu$ is a renormalization and factorization scale. $\gamma=
-\alpha_s/\pi$ is the quark anomalous dimension. The expression of $s$ 
is very complicated and referred to \cite{L1,WYL}.
The initial conditions $\phi_i(x)$, 
$i=B$, $D$ and $D^*$, are of nonperturbative origin, satisfying the 
normalization $\int_0^1\phi_i(x)dx=f_i/(2\sqrt{6})$, $f_i$ being the 
corresponding meson decay constants. 

The evolution of the hard scattering amplitude $H$ is expressed as 
\cite{L1}
\begin{equation}
H(k_1^-,k_2^+,b_1, b_2,\mu)= H(k_1^-,k_2^+,b_1,b_2,t)\exp
\left[-4\int_{\mu}^{t}\frac{d{\bar \mu}}{\bar \mu}\gamma(\alpha_s
({\bar\mu}))\right]\;,
\label{eh}
\end{equation}
where the variable $t$ denotes the largest mass scale of $H$. Combining
Eqs.~(\ref{wp}) and (\ref{eh}), we obtain the factorization formulas
of $\xi_i$. 

For nonleptonic decays, there exist additional important corrections from 
final-state interactions with soft gluons attaching the outgoing hadrons.
It has been argued that these corrections produce only single logarithms
\cite{L1}, and are thus not considered here.
\vskip 1.0cm

\centerline{\large \bf 3. Nonleptonic $B\to D^{(*)}$ decays}
\vskip 0.3cm

In the analysis of nonleptonic $B$ meson decays we employ the Hamiltonian
in Eq.~(\ref{ham}), instead of the effective Hamiltonian in (\ref{eff}). 
We shall address how to formulate our PQCD theory based on the 
effective Hamiltonian in the end of this letter.
The amplitudes for charged and neutral $B$ 
meson decays are then written as
\begin{eqnarray}
\frac{G_F}{\sqrt{2}}V_{cb}V_{ud}^*\left[\langle\pi^-|({\bar d}u)
|0\rangle\langle D^{(*)0}|({\bar c}b)|B^-\rangle-
\langle\pi^-|\langle D^{(*)0}|({\bar d}u)({\bar c}b)|B^-\rangle\right],
\label{ch}\\
\frac{G_F}{\sqrt{2}}V_{cb}V_{ud}^*\left[\langle\pi^-|({\bar d}u)
|0\rangle\langle D^{(*)+}|({\bar c}b)|{\bar B}^0\rangle-
\langle\pi^-|\langle D^{(*)+}|({\bar d}u)({\bar c}b)|{\bar B}^0
\rangle\right].
\label{ne}
\end{eqnarray}
The first terms in Eqs.~(\ref{ch}) and (\ref{ne}) correspond to the external
$W$-emission contributions, and the second terms to
the internal $W$-emission and $W$-exchange contributions, respectively.
Note the minus signs in front of the second terms, which are associated
with the interchange of two $u$ ($d$) quarks in the internal $W$-emission
($W$-exchange) diagrams as shown in Fig.~1(b) (Fig.~1(c)) compared to
Fig.~1(a). There are another two nonfactorizable internal $W$-emission 
diagrams, in which one end of the gluon line attaches the quark of the 
$D^{(*)}$ meson. However, it can be shown that the contributions from 
these two diagrams cancel partially, and are thus less important. For the 
same reason, we neglect the contributions from another two nonfactorizable 
$W$-exchange diagrams, in which the gluon attaches the quark of the $B$ 
meson.

It has been argued that the internal $W$-emission and $W$-exchange 
amplitudes can be evaluated in the PQCD approach \cite{WYL}. 
The decay rates of $B\to D^{(*)}$ transitions have the expression
\begin{equation}
\Gamma_i=\frac{1}{128\pi}G_F^2|V_{cb}|^2|V_{ud}|^2m_B^3\frac{(1-r^2)^3}{r}
|{\cal M}_i|^2\;,
\end{equation}
with $r=M_{D^{(*)}}/M_B$ and $i=1$, 2, 3 and 4 denoting 
$B^-\to D^0\pi^-$, ${\bar B}^0\to D^+\pi^-$, $B^-\to D^{*0}\pi^-$ and
${\bar B}^0\to D^{*+}\pi^-$, respectively. The amplitudes ${\cal M}_i$
are written as
\begin{eqnarray}
{\cal M}_1&=&f_\pi[(1+r)\xi_+-(1-r)\xi_-]+\frac{f_D}{N_c}
\xi_{\rm int}\;,
\label{M1}\\
{\cal M}_2&=&f_\pi[(1+r)\xi_+-(1-r)\xi_-]+\frac{\pi^2}{4N_c}f_B
\xi_{\rm exc}\;,
\label{M2}\\
{\cal M}_3&=&\frac{1+r}{2r}f_\pi[(1+r)\xi_{A_1}-(1-r)(r\xi_{A_2}
+\xi_{A_3})]+\frac{f_{D^*}}{N_c}\xi^*_{\rm int}\;,
\label{M3}\\
{\cal M}_4&=&\frac{1+r}{2r}f_\pi[(1+r)\xi_{A_1}-(1-r)(r\xi_{A_2}
+\xi_{A_3})]+\frac{\pi^2}{4N_c}f_B\xi^*_{\rm exc}\;.
\label{M4}
\end{eqnarray}
The color-suppressing factors $1/N_c$ have been shown explicitly.

The factorization formulas for $\xi_i$, $i=+$, $V$, $A_1$ and $A_3$ 
($\xi_{A_3}=\xi_V$) and for $\xi_j$, $j=-$ and $A_2$ are given by
\cite{WYL}
\begin{eqnarray}
\xi_i&=& 16\pi{\cal C}_F\sqrt{r}M_B^2
\int_{0}^{1}d x_{1}d x_{2}\,\int_{0}^{\infty} b_1d b_1 b_2d b_2\,
\phi_B(x_1)\phi_{D^{(*)}}(x_2)\alpha_s(t)
\nonumber \\
& &\times [(1+\zeta_ix_2r)h(x_1,x_2,b_1,b_2,m)
+(r+\zeta'_ix_1)h(x_2,x_1,b_2,b_1,m)]
\nonumber \\
& &\times \exp[-S(x_1,x_2,b_1,b_2)]\;,
\label{+}\\
\xi_j&=& 16\pi{\cal C}_F\sqrt{r}M_B^2
\int_{0}^{1}d x_{1}d x_{2}\,\int_{0}^{\infty} b_1d b_1 b_2d b_2\,
\phi_B(x_1)\phi_{D^{(*)}}(x_2)\alpha_s(t)
\nonumber \\
& &\times [\zeta_j x_2rh(x_1,x_2,b_1,b_2,m)
+\zeta'_jx_1 h(x_2,x_1,b_2,b_1,m)]
\nonumber \\
& &\times \exp[-S(x_1,x_2,b_1,b_2)]\;,
\label{-}
\end{eqnarray}
with the constants
\begin{eqnarray}
& &\zeta_+=\zeta'_+=\frac{1}{2}\left[\eta-\frac{3}{2}+
\sqrt{\frac{\eta-1}{\eta+1}}\left(\eta-\frac{1}{2}\right)\right]\;,
\nonumber \\
& &\zeta_V=-\frac{1}{2}-\frac{\eta-2}{2\sqrt{\eta^2-1}}\;,
\;\;\;\;\zeta'_V=\frac{1}{2\sqrt{\eta^2-1}}\;,
\nonumber \\
& &\zeta_{A_1}=-\frac{2-\eta-\sqrt{\eta^2-1}}{\eta+1}\;,
\;\;\;\;\zeta'_{A_1}=\frac{1}{2(\eta+1)}\;,
\nonumber \\
& &\zeta_-=-\zeta'_-=-\frac{1}{2}\left[\eta-\frac{1}{2}+
\sqrt{\frac{\eta+1}{\eta-1}}\left(\eta-\frac{3}{2}\right)\right]\;,
\nonumber \\
& &\zeta_{A_2}=0\;,\;\;\;\;\zeta'_{A_2}=-1-\frac{\eta}{\sqrt{\eta^2-1}}\;.
\end{eqnarray}
${\cal C}_F=4/3$ is the color factor. These form factors are evaluated at 
the maximal recoil $\eta=\eta_{\rm max}=(1+r^2)/(2r)$ in
Eqs.~(\ref{M1}) to (\ref{M4}). 

The form factors $\xi^{(*)}_{\rm int}$ and $\xi^{(*)}_{\rm exc}$ are 
given by
\begin{eqnarray}
\xi^{(*)}_{\rm int}&=&16\pi{\cal C}_F\sqrt{r}M_B^2
\int_0^1 dx_1dx_3\int_0^{\infty}b_1db_1b_3db_3
\phi_B(x_1)\phi_\pi(x_3)\alpha_s(t_{\rm int})
\nonumber \\
& &\times \left[(1+x_3(1-r^2))h_{\rm int}(x_1,x_3,b_1,b_3,m_{\rm int})
\right.
\nonumber \\
& &\hskip 0.3cm \left.+\zeta^{(*)}_{\rm int}x_1r^2 
h_{\rm int}(x_3,x_1,b_3,b_1,m_{\rm int})\right]
\exp[-S_{\rm int}(x_1,x_3,b_1,b_3)]\;,
\label{int} \\\
\xi^{(*)}_{\rm exc}&=&16\pi{\cal C}_F\sqrt{r}M_B^2
\int_0^1 dx_2dx_3\int_0^{\infty}b_2db_2b_3db_3
\phi_{D^{(*)}}(x_2)\phi_\pi(x_3)\alpha_s(t_{\rm exc})
\nonumber \\
& &\times\left[(x_3(1-r^2)-\zeta^{(*)}_{\rm exc}r^2)
h_{\rm exc}(x_2,x_3,b_2,b_3,m_{\rm exc})\right.
\nonumber \\
& &\hskip 0.3cm \left.-x_2h_{\rm exc}(x_3,x_2,b_3,b_2,m_{\rm exc})\right]
\exp[-S_{\rm exc}(x_2,x_3,b_2,b_3)]\;,
\label{exc}
\end{eqnarray}
with the constants $\zeta_{\rm int}=-\zeta^*_{\rm int}=1$ and
$\zeta_{\rm exc}=-\zeta^*_{\rm exc}=1$. In the derivation of 
$\xi_{\rm int}^{(*)}$ we have assumed that $k_1$ has a large plus 
component $k_1^+=x_1P_1^+$. Here $x_3$ is the momentum 
fraction associated with the pion, and $b_3$ can be regarded as 
the spatial extent of the pion. $\xi^{(*)}_{\rm int}$ and 
$\xi^{(*)}_{\rm exc}$ are in fact the $B\to\pi$ and $D\to\pi$ transition 
form factors, respectively, evaluated at $\eta=\eta_{\rm max}$. 

The Sudakov exponents $S$'s, which group the exponents in Eqs.~(\ref{wp}) 
and (\ref{eh}), are given by
\begin{eqnarray}
S&=&s(x_1P_1^-,b_1)+s(x_2P_2^+,b_2)+s((1-x_2)P_2^+,b_2)
\nonumber \\
& &-\frac{1}{\beta_1}\left[\ln\frac{\ln(t/\Lambda)}{-\ln(b_1\Lambda)}
+\ln\frac{\ln(t/\Lambda)}{-\ln(b_2\Lambda)}\right]
\label{S}\\
S_{\rm int}&=&s(x_1P_1^+,b_1)+s(x_3P_3^-,b_3)+
s((1-x_3)P_3^-,b_3)
\nonumber \\
& &-\frac{1}{\beta_1}\left[\ln\frac{\ln(t_{\rm int}/\Lambda)}
{-\ln(b_1\Lambda)}+\ln\frac{\ln(t_{\rm int}/\Lambda)}
{-\ln(b_3\Lambda)}\right]\;,
\\
S_{\rm exc}&=&s(x_2P_2^+,b_2)+s((1-x_2)P_2^+,b_2)+s(x_3P_3^-,b_3)
\nonumber \\
& &+s((1-x_3)P_3^-,b_3)
-\frac{1}{\beta_1}\left[\ln\frac{\ln(t_{\rm exc}/\Lambda)}
{-\ln(b_2\Lambda)}+\ln\frac{\ln(t_{\rm exc}/\Lambda)}
{-\ln(b_3\Lambda)}\right]\;,
\end{eqnarray}
with $\beta_1=(33-2n_f)/12$ and $n_f=4$ the number of flavors. The QCD 
scale $\Lambda\equiv\Lambda_{\rm QCD}$ will be set to 0.2 GeV below. 
The factors $e^{-S}$ fall off quickly in the large $b$, or 
long-distance, region, giving so-called Sudakov suppression.

In Eqs.~(\ref{+}), (\ref{-}), (\ref{int}) and (\ref{exc}) the functions 
$h$'s, obtained from the Fourier transform of the lowest-order $H$, are 
given by
\begin{eqnarray}
h(x_1,x_2,b_1,b_2,m)&=&K_{0}\left(\sqrt{x_1x_2m}b_1\right)
\nonumber \\
& &\times \left[\theta(b_1-b_2)K_0\left(\sqrt{x_2m}
b_1\right)I_0\left(\sqrt{x_2m}b_2\right)\right.
\nonumber \\
& &\;\;\;\;\left.
+\theta(b_2-b_1)K_0\left(\sqrt{x_2m}b_2\right)
I_0\left(\sqrt{x_2m}b_1\right)\right]\;,
\label{dh}\\
h_{\rm int}(x_1,x_3,b_1,b_3,m_{\rm int})&=&h(x_1,x_3,b_1,b_3,
m_{\rm int})\;,
\\
h_{\rm exc}(x_2,x_3,b_2,b_3,m_{\rm exc})&=&
H_0^{(1)}\left(\sqrt{x_2x_3m_{\rm exc}}b_2\right)
\nonumber \\
& &\times\left[\theta(b_2-b_3)
H_0^{(1)}\left(\sqrt{x_3m_{\rm exc}}b_2\right)
J_0\left(\sqrt{x_3m_{\rm exc}}b_3\right)\right.
\nonumber \\
& &\left.+\theta(b_3-b_2)H_0^{(1)}\left(\sqrt{x_3m_{\rm exc}}b_3\right)
J_0\left(\sqrt{x_3m_{\rm exc}}b_2\right)\right]\;,
\nonumber \\
& &
\end{eqnarray}
with $m=(\eta+\sqrt{\eta^2-1})M_BM_{D^{(*)}}$
and $m_{\rm int}=m_{\rm exc}=M_B^2-M_{D^{(*)}}^2$.
It is obvious that the $W$-exchange contribution is complex due 
to the exchange of a time-like hard gluon. 
In the above expressions the large scales $t$'s take the values
\begin{eqnarray}
t&=&{\rm max}(\sqrt{x_1x_2m},1/b_1,1/b_2) \\
t_{\rm int}&=&{\rm max}(\sqrt{x_1x_3m_{\rm int}},1/b_1,1/b_3) \\
t_{\rm exc}&=&{\rm max}(\sqrt{x_2x_3m_{\rm exc}},1/b_2,1/b_3)\;. 
\end{eqnarray}

We choose $f_B=200$ MeV, $f_D=f_{D^*}=220$ MeV \cite{A}, $|V_{cb}|=0.043$ 
for the CKM matrix element \cite{L1,WYL}, and the Chernyak-Zhitnitsky 
model \cite{CZ}
\begin{equation}
\phi_\pi(x)=\frac{5\sqrt{6}}{2}f_\pi x(1-x)(1-2x)^2
\end{equation}
for the pion wave function,
$f_\pi=132$ MeV being the pion decay constant.
For the $B$ meson, we employ the wave function from the relativistic 
constituent quark model \cite{S}, 
\begin{equation}
\phi_B(x,{\bf k}_T)=N_B\left[C_B+\frac{M_B^2}{1-x}+\frac{k_T^2}{x(1-x)}
\right]^{-2}\;.
\end{equation}
The normalization constant $N_B$ and the shape parameter $C_B$ are 
determined by the conditions
\begin{eqnarray}
& &\int_0^1dx\int \frac{d^2{\bf k}_T}{16\pi^3} \phi_B(x,{\bf k}_T)
=\frac{f_B}{2\sqrt{6}}\;,
\nonumber \\
& &\int_0^1dx\int \frac{d^2{\bf k}_T}{16\pi^3} [\phi_B(x,{\bf k}_T)]^2=
\frac{1}{2}\;.
\label{cs}
\end{eqnarray}
The $B$ meson wave function is then given by
\begin{equation}
\phi_B(x)=\int \frac{d^2{\bf k}_T}{16\pi^3} \phi_B(x,{\bf k}_T)
=\frac{N_B}{16\pi^2}\frac{x(1-x)^2}{M_B^2+C_B(1-x)}\;,
\label{bw}
\end{equation}
with $N_B=650.212$ and $C_B=-27.1051$.
We assume that the $D^{(*)}$ meson wave function possesses
the same functional form as Eq.~(\ref{bw}),
\begin{equation}
\phi_{D^{(*)}}(x)=\frac{N_{D^{(*)}}}{16\pi^2}\frac{x(1-x)^2}
{M_{D^{(*)}}^2+C_{D^{(*)}}(1-x)}\;,
\label{dw}
\end{equation}
but the shape parameter $C_D^{(*)}$ is fixed by the $B\to D^{(*)}$ data
\cite{A}. 
\vskip 1.0cm

\centerline{\large \bf 4. Discussion}
\vskip 0.3cm

We adopt $G_F=1.16639\times 10^{-5}$ GeV$^{-2}$, $|V_{ud}|=0.974$, 
$M_B=5.28$ GeV, $M_D=1.87$ GeV, $M_{D^*}=2.01$ GeV \cite{PDG}, and
$\tau_{B^0}=1.53$ ($\tau_{B^-}=1.68$) ps for the ${\bar B}^0$ 
($B^-$) meson lifetime \cite{B}. The experimental data of the branching 
ratios are
${\cal B}(B^-\to D^0\pi^-)=(5.5\pm 1.1)\times 10^{-3}$,
${\cal B}({\bar B}^0\to D^+\pi^-)=(2.9\pm 1.2)\times 10^{-3}$,
${\cal B}(B^-\to D^{*0}\pi^-)=(5.2\pm 1.7)\times 10^{-3}$, and
${\cal B}({\bar B}^0\to D^{*+}\pi^-)=(2.6\pm 0.8)\times 10^{-3}$ \cite{A}.
We determine the parameter $C_D^{(*)}$, {\it ie.}, the $D^{(*)}$ meson
wave function, by fitting our predictions to 
${\cal B}({\bar B}^0\to D^+\pi^-)$ and
${\cal B}({\bar B}^0\to D^{*+}\pi^-)$. They are found to be
$C_D=-2.95$ GeV$^2$ and $C_{D^*}=-3.05$ GeV$^2$. The corresponding 
normalization constants are then $N_D=128.77$ GeV$^3$ and $N_{D^*}=173.48$
GeV$^3$.

With these parameters we derive the branching ratios 
${\cal B}(B^-\to D^0\pi^-)=4.41\times 10^{-3}$, 
${\cal B}({\bar B}^0\to D^+\pi^-)=2.91\times 10^{-3}$,
${\cal B}(B^-\to D^{*0}\pi^-)=3.97\times 10^{-3}$, and
${\cal B}({\bar B}^0\to D^{*+}\pi^-)=2.59\times 10^{-3}$, which are 
within the errors of the data. It is observed that the internal $W$-emission
amplitude is additive to and about 20\% of the external $W$-emission
amplitude. Therefore, our analysis favors a positive $a_2/a_1$ for $B$ 
meson decays. The internal $W$-emission
contribution obtained in \cite{CM} is complex, but it is real in our
formalism, and is twice larger in magnitude. Hence, our predictions
for the charged $B$ meson decays are in a better agreement with the data.
We also find that the $W$-exchange amplitude is mainly imaginary, and its 
magnitude is only 3\% of the external $W$-emission amplitude. The real 
part of the $W$-exchange contribution is at least one order of magnitude 
smaller than the imaginary part. The corresponding results in \cite{CM} 
have the same order, but the real part is larger or only slightly smaller 
than the imaginary part in magnitude for different models of pion wave 
functions. The branching ratios of the charged $B$ meson decays are then
$(1.2)^2\times (\tau_{B^-}/\tau_{B^0})\approx 1.5$ times of
those of the neutral $B$ meson decays, fairly consistent with the data,
and with the prediction ${\cal B}(B^-\to D^{(*)0}\pi^-)/
{\cal B}({\bar B}^0\to D^{(*)+}\pi^-)\approx 1.7$ from the BSW method 
\cite{A}.

The branching ratios of the charged $B$ meson decays fall below 
the central values of the data. The agreement can be improved by including
the two nonfactorizable internal $W$-emission diagrams. Because of the 
complexity of their evaluation, we shall not study these two diagrams in 
this letter, but discuss them elsewhere. 

It is worthwhile to exhibit the spectrum $d\Gamma/dq^2$ 
of the semileptonic decay ${\bar B}^0 \to D^{*+}\ell^-{\bar \nu}$. 
For this process only the external $W$-emission diagrams 
contribute, and the expression is written as \cite{WYL}
\begin{eqnarray}
\frac{d \Gamma}{d q^2}&=&
\frac{1}{96\pi^3}G_F^2|V_{cb}|^2M_B^3r^2(\eta^2-1)^{1/2}(\eta+1)^2
\nonumber\\
& &\times\left\{2(1-2\eta r+r^2)\left[\xi_{A_1}^2(\eta)+\frac{\eta-1}
{\eta+1}\xi_V^2(\eta)\right]\right.
\nonumber \\
& &\hskip 0.5cm +\left[(\eta-r)\xi_{A_1}(\eta)-(\eta-1)
\left(r\xi_{A_2}(\eta)+\xi_{A_3}(\eta)\right)\right]^2\Biggr\}\;,
\label{dd}
\end{eqnarray}
where the form factors $\xi_i(\eta)$ have been defined in Eqs.(\ref{+}) 
and (\ref{-}). It is observed from Fig.~2 that our predictions match
the data \cite{B} at low $q^2$, but begin to deviate above $q^2=4$ GeV$^2$, 
the slow recoil region in which PQCD is not reliable. 

Since the $W$-exchange contributions are negligible, and the $B\to \pi$ 
and $B\to \rho$ form factors are roughly equal \cite{LY1}, the branching
ratios of the decays $B\to D^{(*)}\rho$ can be easily estimated by 
substituting the $\rho$ meson decay constant $f_\rho=220$ MeV for $f_\pi$ 
in the corresponding formulas. Assuming a vanishing $\rho$ meson mass, 
we have ${\cal B}(B^-\to D^0\rho^-)=1.2\%$, 
${\cal B}({\bar B}^0\to D^+\rho^-)=8.1\times 10^{-3}$, 
${\cal B}(B^-\to D^{*0}\rho^-)=1.1\%$,
and ${\cal B}({\bar B}^0\to D^{*+}\rho^-)=7.2\times 10^{-3}$. 
Compared to the data $(1.35\pm 0.30)\%$, $(8.1\pm 3.6)\times 10^{-3}$, 
$(1.68\pm 0.58)\%$, and $(7.4\pm 2.7)\times 10^{-3}$ \cite{A}, 
respectively. Our predictions are satisfactory except for the mode
$B^-\to D^{*0}\rho^-$. Similarly, the overall consistency can be 
improved by including the two nonfactorizable internal $W$-emission 
diagrams.

It is nontrivial to incorporate the effective Hamiltonian in 
Eq.~(\ref{eff}) into factorization theorems. 
The Wislson coefficients $c_1$ and $c_2$, or $a_1$ and $a_2$ equivalently,
which take into account the evolution from the $W$ boson mass $M_W$
down to the scale $\mu$, arise from the irreducible radiative
corrections to the decay vertex. In the conventional approach $\mu$ 
is set to a value of order $M_b$, and thus a scale dependence is 
introduced. We argue that $\mu$ in $a_1$ and $a_2$ should be set to the 
large scale $t$ of the hard scattering. The scale $t$ is then integrated 
over, such that predictions remain RG invariant. Hence, $a_1$ and $a_2$ 
are not parameters as in the BSW model, but vary
according to their evolutions in our approach. 
The controversy over the scale setting in the analysis of inclusive 
nonleptonic $B$ meson decays \cite{LSW} may be resolved by our RG 
invariant formalism. 

An immediate observation from the above choice of $\mu$ is that the 
internal $W$-emission amplitudes are still positive in bottom decays. 
If main contributions came from the region with $t$ close to the $b$ quark 
mass $M_b$, the coefficients of the external and internal $W$-emission 
amplitudes, $a_1=1$ and $a_2=1/N_c$, are replaced by $a_1(M_b)=1.03$ and 
$a_2(M_b)=0.11$. The internal $W$-emission amplitudes, however,
may become negative for charm decays because of $a_1(M_c)=1.09$ and 
$a_2(M_c)=-0.09$, $M_c$ being the $c$ quark mass. Certainly, this issue 
on the sign of $a_2/a_1$ \cite{BSW,KP} needs more quantitative study.
The details will be published in a separate work.

Effects from nonfactorizable final-state interactions \cite{CT}
can be included into our formalism easily, which will be 
discussed elsewhere. The discrepancy between 
model-dependent theoretical predictions and experimental data associated 
with the decays $B\to\psi K^{(*)}$ \cite{GKP} can also be investigated.
\vskip 0.5cm

We thank H.Y. Cheng for useful discussions.
This work was supported by National Science council of R.O.C. under
the Grant No. NSC-85-2112-M-194-009.

\newpage
\centerline{\large \bf Figure Captions}
\vskip 0.5cm

\noindent
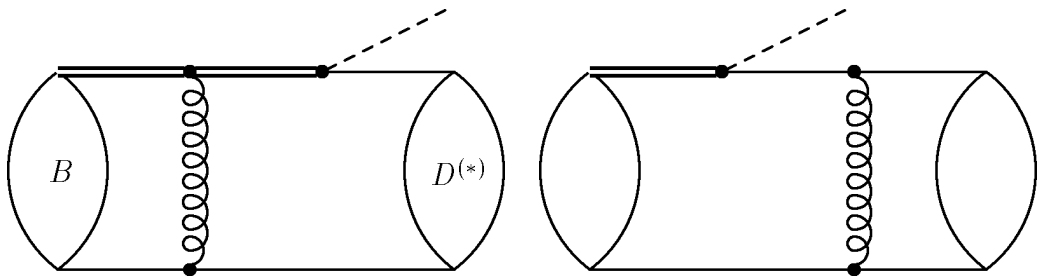
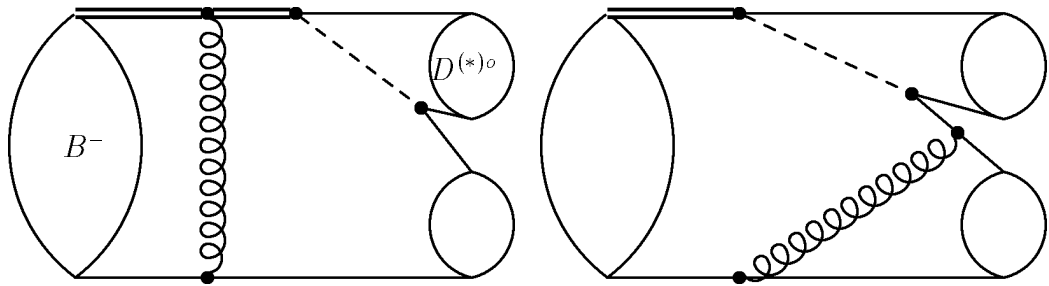
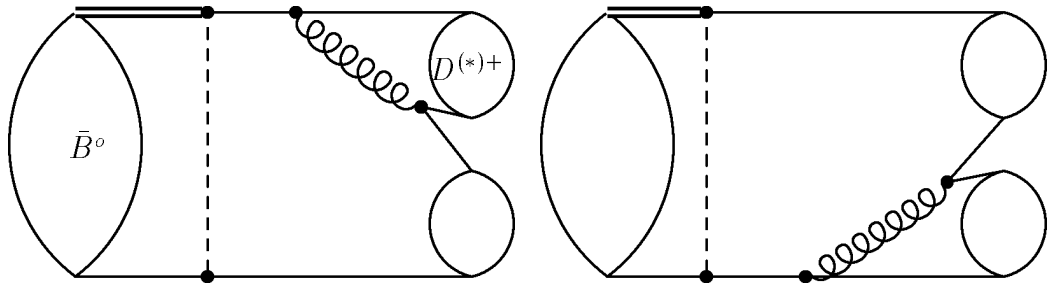
{\bf Fig. 1.} (a) External $W$-emission, (b) internal $W$-emission, and
(c) $W$-exchange diagrams with the $b$ quarks and the $W$ bosons 
represented by double lines and dashed lines, respectively.
\vskip 0.5cm

\noindent
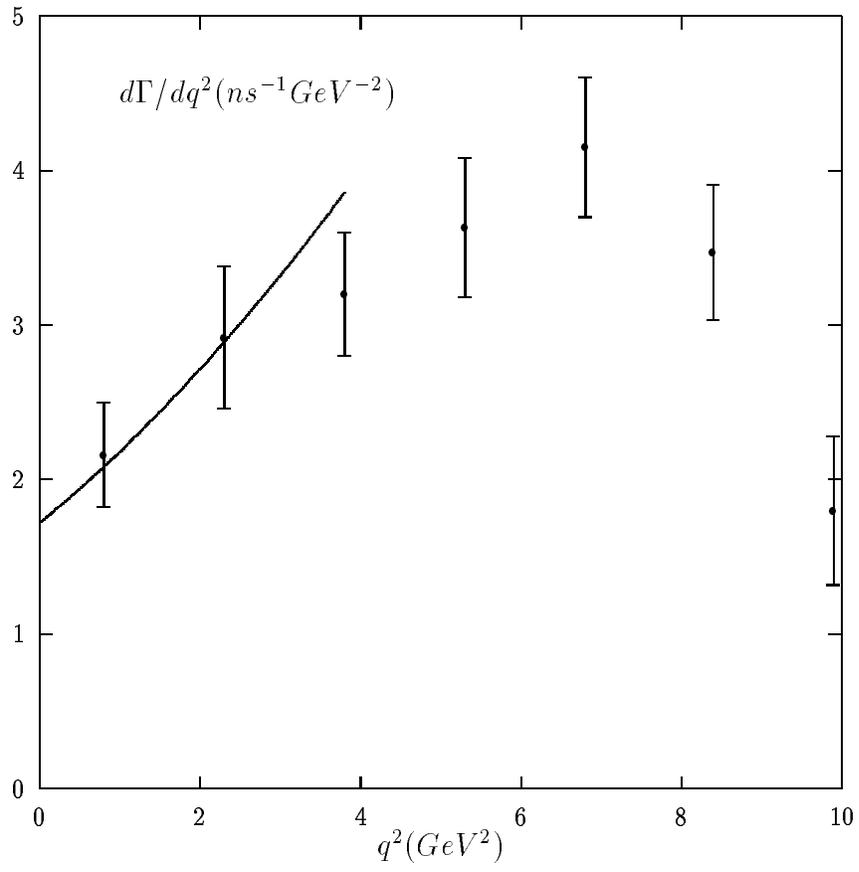
{\bf Fig. 2.} The spectrum $d\Gamma/dq^2$ of the semileptonic decay 
${\bar B}^0\to D^{*+}\ell^-{\bar \nu}$.
\vskip 0.5cm

\end{document}